\documentclass[12pt,preprint]{aastex}
\shorttitle{Lensed Image Angles and Substructure}
\shortauthors{Williams et al.}

\begin{document}

\title{Lensed Image Angles: New Statistical Evidence for Substructure}

       \author{Liliya L.R. Williams, Patrick Foley, Damon Farnsworth}
       \affil{Department of Astronomy\\
              University of Minnesota\\
              116 Church Street SE\\
              Minneapolis, MN 55455}
       \email{llrw@astro.umn.edu}
       \and

       \author{Jason Belter}
       \affil{Nova Classical Academy\\
              1668 Montreal Ave\\
              St. Paul, MN 55116}

\begin{abstract}
We introduce a novel statistical way of analyzing the projected mass distribution 
in galaxy lenses based solely on the angular distribution of images in quads
around the lens center. The method requires the knowledge of the lens center
location, but the images' distances from the lens center are not
used at all. If the images of a quad are numbered in order of arrival time, 
$\theta_1$ through $\theta_4$, and $\theta_{ij}$ is the angle between images 
$i$ and $j$, then we define the 'bisector' plane whose axes are linear 
combinations of $\theta_{23}$ and $\theta_{14}$. The bisector plane of a 
given lens contains all the quads produced by the lens. We show empirically 
that all two-fold symmetric lenses with convex, i.e. non-wavy or petal-like 
isodensity contours are identical in the bisector plane of their quads. 
We also study lenses with twisting isodensity contours, lumpy substructure, 
etc. Our results suggest that to reproduce the general characteristics of 
the observed quad population, kpc-scale substructure must be a common feature 
of galaxy lenses.
\end{abstract}


\section{Introduction}\label{intro}

In the last decade or so gravitationally lensed QSOs, both doubles and 
quads, have been used mostly for the determination of the Hubble parameter 
(see \citet{coles08} for the latest work, and summary of earlier results), 
and for the estimation of the mass distribution in the lensing galaxies. 
In this paper we will concentrate on the latter.

One can loosely divide the information on the lens mass distribution into
two categories: radial and angular. Much attention has been paid in the 
literature to the sky-projected radial mass distribution 
in lenses because the slope of the density profile, and its variation with
radius is a test of cosmological models \citep{nfw96,nfw97}.
The density profile slope in the 
central regions is also important because it is affected by the (adiabatic) 
contraction of dark matter halos in response to the collapsing baryons during 
galaxy formation \citep{fsw05,fsb08,g08}.

The angular distribution of lensing mass, for example, the degree of
ellipticity, the change in the ellipticity position angle with radius,
etc. have received some attention as well \citep{cdk08,sw06,ok04}, but mostly
as ``nuisance'' parameters in determining the radial density profile or the 
Hubble constant.  It is somewhat ironic that the
generally uninteresting ellipticity position angle can be unambiguously 
estimated by any reasonable modeling method, even by eye \citep{sw03},
whereas, the more interesting density profile slope is often very 
uncertain because of the mass-sheet, or steepness degeneracy \citep{fgs85,s00}. 

The positions of lensed images of a quad or a double can also be looked
at as consisting of angular and radial information. By radial information 
we mean the relative spread of images in distance from the lens center. 
The angular information is the angular separation of the images as viewed 
from the lens center. For example, in the Cloverleaf, H1413+117, and
the Einstein Cross, Q2237+030 any two adjacent images are roughly $90^\circ$
apart. In doubles, the two non-central images tend to be separated by 
$\sim 150^\circ-180^\circ$.

Obviously there is no simple one-to-one relation between, say, the radial 
structure of the lensing mass and the radial distribution of lensed images.  
However, there are some qualitative connections between the two. 
For example, a steep lens mass distribution tends to produce quads
with narrow radial spread of images, largely independent of the angular 
distribution of these images, or the ellipticity of the lensing mass. 
Conversely, if the lensing mass has a shallow density profile the images 
tend to have a wider radial spread. In the Appendix of this paper we show 
that one angular property of the lensing mass, its ellipticity position 
angle can be straightforwardly and rather precisely estimated from the 
angular positions of the four images of the quad (Section~\ref{estimatingPA}). 

The main work presented in this paper is loosely motivated by the 
preceding paragraph. Specifically, we ask what information about the
lensing mass can be retrieved by looking solely at the angular distribution
of lensed images around the lens center.

\section{Defining angles and bisector rays}\label{defining}

\begin{figure}
\epsscale{0.9}
\plotone{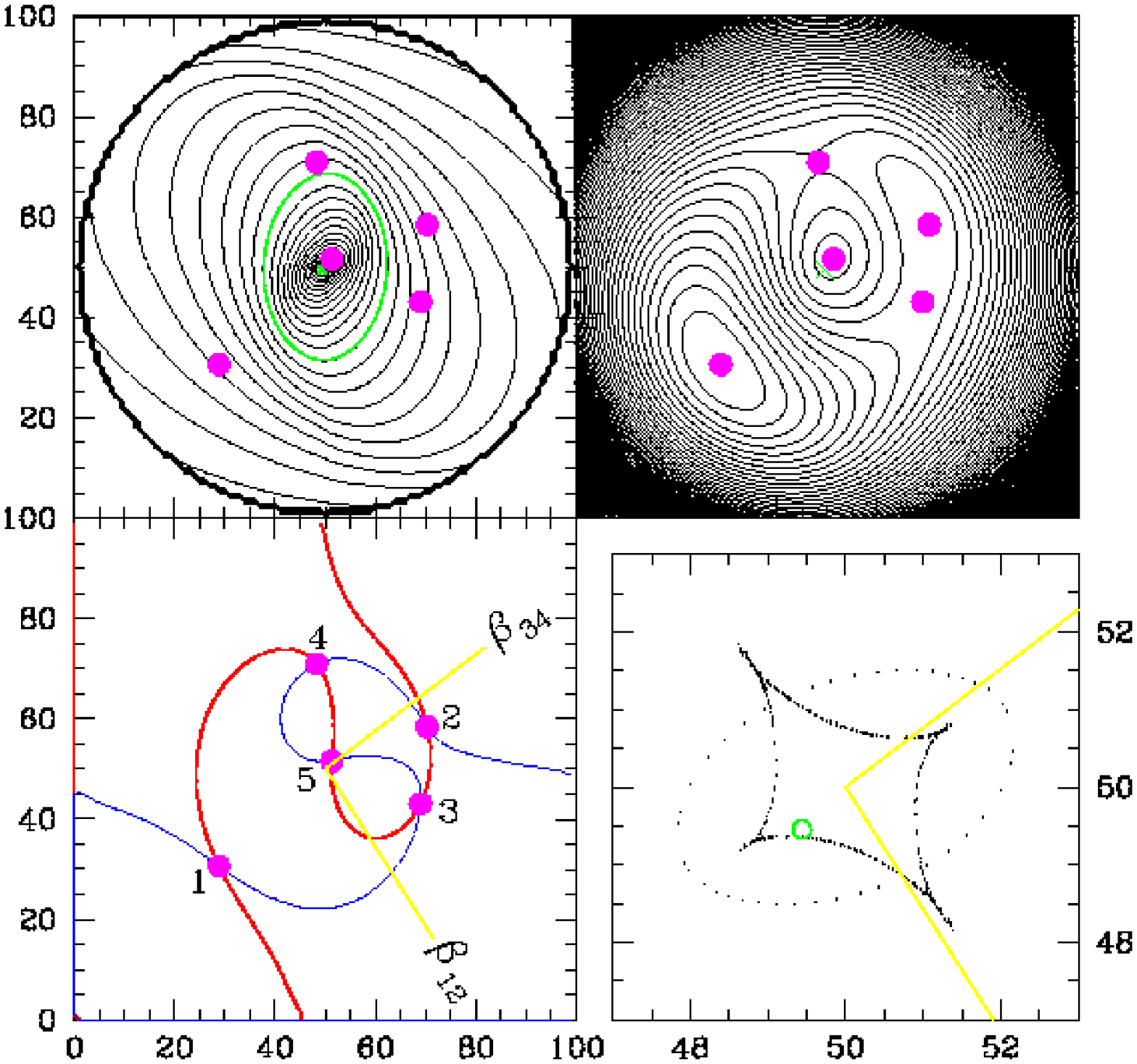}
\caption{A synthetic galaxy lens. 
{\em Upper left:} Contours of surface mass density, isodens, of the mass 
distribution of the lens. The thick (green) contour marks the critical 
lensing surface mass density. Images are filled (magenta) dots. The 
contours are spaced linearly. The mass was defined in a circular window. 
{\em Upper right:} Arrival time surface, with images.
{\em Lower left:} The thick (red) and thin (blue) lines are the solutions 
of the lens equation in the $x$- and $y$-directions, respectively. 
The intersections are the image positions. Images are labeled according 
to the arrival time, from 1 to 5. Two bisector rays, $\beta_{12}$ and 
$\beta_{34}$ are drawn as solid (yellow) lines.
{\em Lower right:} Source plane caustics. The straight lines are the bisector 
rays. The position of the source is marked with an empty (green) circle.}
\label{fourpanels}
\end{figure}

Following \citet{sw03}, we refer to the four images of a quad by their 
arrival time, as 1,~2,~3,~4. Image 1 is the global minimum of the arrival
time surface and hence is the first arriving image. Image 2 is the second 
arriving image, and is also a minimum. Images 3 and 4 are saddles of the 
arrival time surface. Image 5, a maximum, is the central demagnified image,
and is usually not detected. (See Figure~\ref{fourpanels}). As explained in 
\citet{sw03} figuring out the arrival {\it order} of images in observed quads
can be done, in most cases, based on the morphology of the image distribution 
alone, without measuring the time delays. 

Images 2 and 3 (minimum and saddle) often appear close together; these are 
the two images that merge and disappear when the source moves away from the 
lens center. Because of that, the angular separation of these two images 
(as seen from the lens center), which we will call $\theta_{23}$ can be a 
measure of the "quadrupoleness" of a quad system. When 2 and 3 are close 
together the system is barely a quad, and could have been a double if the 
source happened to be somewhat further away from the lens center, whereas 
a quad with images 2 and 3 about $90^\circ$ apart is a ``well established'' 
quad. 

We also define $\beta_{12}$, as the ray anchored at the lens center that bisects 
the angle between images 1 and 2. If we further specify that $\beta_{12}$ 
points roughly away from image 4, then the definition of $\beta_{12}$ is 
unambiguous.
Similarly, we define $\beta_{34}$ as the ray bisecting the angle between 
images 3 and 4, and pointing roughly away from image 1. The two lower panels
in Figure~\ref{fourpanels} show both these rays for a synthetic mass distribution,
whose projected density contours are shown in the upper left panel. 
The images are filled circles. The arrival time surface
is shown in the upper right. The lower left panel shows that the images are
found as the intersection of the solution of the lens equation in the $x$ and 
$y$ directions, shown by thick (red) and thin (blue) curves, respectively. 
The lower right panel shows the source plane caustics, the source position
(empty green circle), and the two bisector rays.

These angles and bisector rays turn out to have some very interesting
properties, which relate to certain aspects of the lens mass distributions.

\section{Mass distribution: Lenses with two-fold symmetry}\label{twofold}

\subsection{Defining two-fold symmetric lenses}\label{deftwofold}

A two-fold symmetric lens is a projected mass distribution that has
two orthogonal axes of bilateral symmetry. A wide class of popular lens models 
are two-fold symmetric. For example, this category includes elliptical 
lenses, with any radial density profile. The degree of ellipticity can be a 
function of radius, but the ellipticity position angle (PA) should not change 
with radius. Lenses with single or multiple external shear axes, as long as 
the shear axes are arranged so as to obey the symmetry, also belong in this 
category. Two lens classes commonly used for parametric modeling, Pseudo 
Isothermal Elliptical Mass Distributions (PIEMD) and Pseudo Isothermal 
Elliptical Potentials (PIEP) \citep{kk93} are also members of the two-fold 
symmetric family of lenses. 

We exclude lenses that, even though two-fold symmetric, have 'wavy' 
isodens. (Isodens are contours of equal projected surface mass density in the lens.)
For example, lenses whose isodens follow $\cos(2n\theta)$, with $n>1$, or where 
isodens look like petals.  In other words, mass distributions with non-convex
isodens are excluded. This is further discussed in Section~\ref{invariant}.
The mass distributions thus defined will be referred to as two-fold symmetric. 

In this paper we examine mass distributions through the properties of the 
quad lenses they generate. Our study is statistical in nature; we use the 
properties of the entire quad population produced by a given mass distribution. 
Insights gained from this study help to draw conclusions from the real data, where
a given galaxy lenses one, or maybe a small handful of sources. 
  
In this Section we discuss two-fold symmetric lenses and show that 
members of this family are indistinguishable when viewed in a diagnostic 
plane whose axes are certain combinations of image angles. Next, we discuss
this diagnostic 'bisector' plane.

\subsection{Introducing the bisector plot}\label{bisector}

The lower right panel of Figure~\ref{fourpanels} suggests that the axes 
containing $\beta_{12}$ and $\beta_{34}$ are good indicators of the 
orientation of the diamond caustic, and by extension, the PA of the major 
and minor axes of the lensing mass distribution around the image ring. 
This statement is quantified in the Appendix; here we use this observation 
to motivate our choice of $\beta_{12}-\beta_{34}$ as an angle that contains
useful information about the lensing mass. 

\begin{figure}
\epsscale{0.8}
\plotone{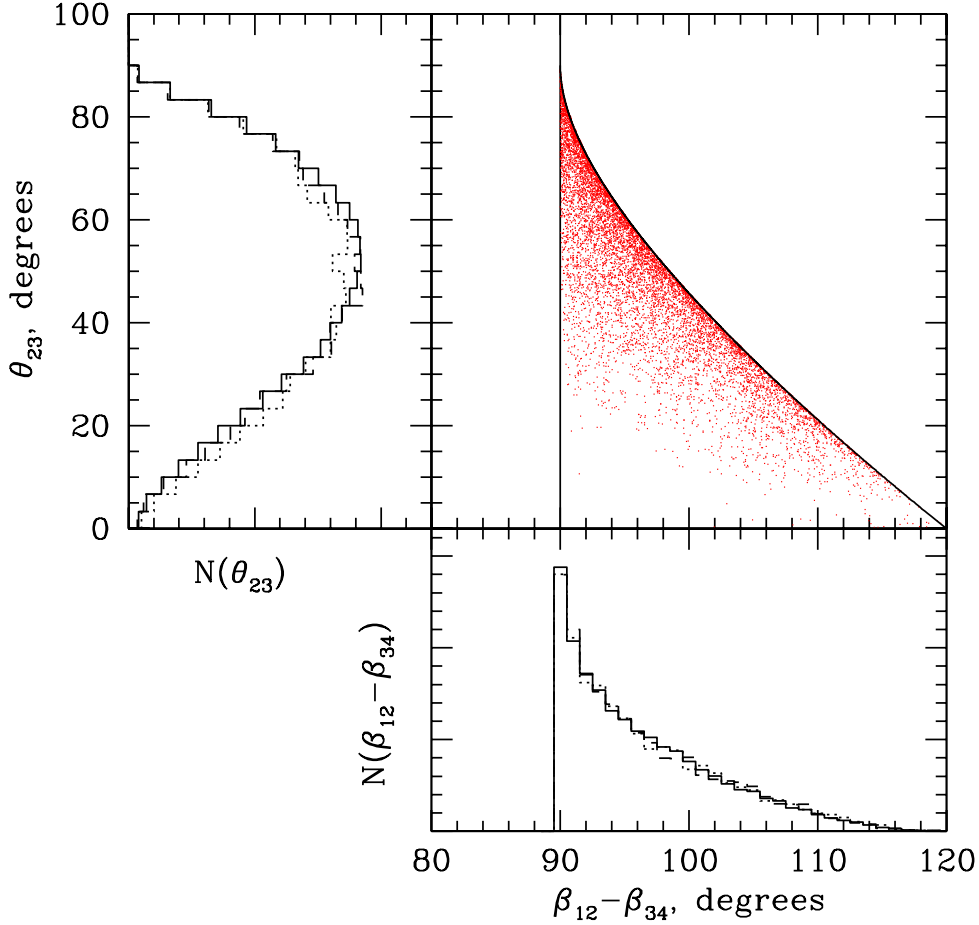}
\caption{The bisector plot for two-fold symmetric lenses as defined in
Section~\ref{deftwofold}:
the difference in bisector angles, $\beta_{12}\!-\!\beta_{34}$ vs. 
the angular separation of images 2 and 3, $\theta_{23}$. 
Each one of the small (red) points corresponds to a different source 
location. The pattern of these points, including the upper envelope, 
appears to be the same for all lenses with two-fold symmetry. 
The solid curve outlining the envelope is given by eqs.~\ref{bisd} and \ref{th23}.
The left and bottom side panels show the distribution of $\theta_{23}$
and $\beta_{12}\!-\!\beta_{34}$ respectively. 
Three different lens models are plotted in the side panels; dashed lines
represent a galaxy lens with a shallow non-power law density profile 
and constant ellipticity of 0.14 (axis ratio 0.75); dotted lines 
represent an ``isothermal'' profile, $\propto R^{-1}$ with a small core, 
and ellipticity of 0.12 (axis ratio 0.79); solid lines represent a 
circular mass distribution with two external axes of shear, $60^\circ$ 
apart, and with shear $\gamma=0.1$. }
\label{bisector_twofold}
\end{figure}

In the main portion of Figure~\ref{bisector_twofold} (upper right panel) 
we plot $\beta_{12}-\beta_{34}$ vs. $\theta_{23}$. Each (red) dot represents a 
4-image lens configuration (a quad); all the dots arise from the same galaxy, 
but each dot has a different source position, picked randomly on the source 
plane. (Sources that do not produce quads did not make it into this plot.) 
The galaxy lens used here has an ``isothermal'' 
projected density profile $\Sigma(R)\propto R^{-1}$ with a small core to avoid 
central singularity. The ellipticity, $\epsilon=0.2$, is constant with radius. 
(The relation between $\epsilon$ and the axis ratio, $r$ is, 
$\epsilon=[1-r]/[1+r]$.)

We call the distribution of points in the $\beta_{12}-\beta_{34}$ vs. 
$\theta_{23}$ plane, the bisector plot. The first thing to note is that
the distribution of points in the bisector plot is not random. 
There are no quads with the bisector difference less 
than $90^\circ$. More interestingly, there is a well defined envelope, a 
curved line above and to the right of which there are no quads. We will 
call this the `envelope'.

The bisector plot of Figure~\ref{bisector_twofold} is 
flanked by two panels. The solid line histogram in the left side panel shows the 
distribution of bisector plot points along the $\theta_{23}$ direction; the 
$\beta_{12}-\beta_{34}$ values have been ``marginalized'' over. The solid line 
histogram in the bottom panel is the distribution of $\beta_{12}-\beta_{34}$ 
values; here, the $\theta_{23}$ values have been marginalized over. These two 
histograms do not fully quantify the distribution of points in the main 
two-dimensional bisector plot, but they do give us an easy, though incomplete 
way of examining that distribution. 
As an example consider a hypothetical quad lens at ($100^\circ,60^\circ$). 
When projected on to the two histograms the point falls in the middle of
both the distributions. So, if one is to ask if this point could have been
drawn from the two distributions, the answer would be 'yes' in both cases.
However, looking at the full 2-d bisector plane it is obvious that the quad
cannot be generated by this lens, as it lies above the bounding envelope,
well outside the distribution.

\subsection{The bisector plot: an invariant property?}\label{invariant}

In the previous section we looked at the bisector plot of one type of lens,
with a certain density profile and certain ellipticity. We have also generated 
bisector plots for many types of lenses, with varying density profiles, 
varying degrees of ellipticity, including ellipticity $\epsilon(r)$ which 
changes in radius, lenses with and without external shear, etc. 
Our numerous experiments suggest that {\it all lenses that possess two-fold 
symmetry, regardless of the radial density distribution and the magnitude or
radial dependence of ellipticity and external shear generate the same 
distribution of points in the bisector plot, bounded by a vertical line 
and a concave envelope.}  We conclude that all two-fold symmetric lenses, as 
defined in Section~\ref{deftwofold} are indistinguishable
in the bisector plot. This is one of the main findings of this paper.

This invariance must derive from the shape of the caustic in the source plane. 
From our experiments we have noticed that the inner (five image) caustics of 
all two-fold symmetric lenses are diamond-shaped, and appear to share 
the following two features. First, the diamond caustic itself has two-fold
symmetry (and so the two lines connecting the opposite cusps are perpendicular 
to each other), and second, the diamond caustics of any two such lenses can be 
made to have the same shape if one is allowed to linearly stretch or shrink 
them in the directions along the lines connecting the opposite cusps. 
By symmetry arguments, the first feature seems natural for lens mass 
distributions that have two-fold symmetry. The lines connecting opposite 
cusps of the diamond caustic of a lens with no such symmetry, for
example the one shown in Figure~\ref{fourpanels} (lower right panel), 
do not intersect at right angles. The second feature implies the invariance
of the caustic itself (modulo linear stretching of the $x$ or $y$ coordinates), 
and is probably the crux of the bisector plot  invariance shown in
Figure~\ref{bisector_twofold}.
 
The invariance does not extend to lenses that have 'wavy' isodens; such lenses 
tend to produce caustics more complicated than diamond shapes.

The invariance does not apply to lenses with naked cusps, i.e. 
lenses whose diamond caustic cusps stick outside of the oval caustic because 
of large ellipticity in the mass distribution.

\subsection{The bisector plot envelope for a specific lensing potential}\label{SISell}

The set of quads that delineate the upper bounding envelope of the bisector 
plane, shown, for example in Figure~\ref{bisector_twofold}, must correspond 
to a continuous set of sources in the source plane of any two-fold 
symmetric lens. We speculate, and confirm using experiments with synthetic 
lenses, that the envelope quads, when mapped back to the source plane, form 
a straight line that connects the center of the lens to the point on the 
diamond caustic closest to the center; we call this the point of closest 
approach, and denote it $\vec r_c$.

If the bisector plane is indeed universal, as we claim, then the envelope
must be described by a universal analytical expression. Here we derive the 
equation for the envelope for a specific type of a two-fold symmetric lens. 

We start with a lensing potential of the form, $\phi(r,\theta)=r\,f(\theta)$
\citep{wmk00}, and work in cylindrical coordinates on the plane of the sky.
The arrival time surface is,
$\psi(r,\theta)=\frac{1}{2}|\vec r-\vec r_s|^2-\phi(r,\theta)$.
The lensing equation, $\vec\nabla\psi=0$, in the $\hat r$ and $\hat \theta$ 
directions is written as,
\begin{equation}
r_s\cos(\theta-\theta_s)=r-f,\quad\quad\quad\quad
r_s\sin(\theta-\theta_s)={{\partial f}\over{\partial\theta}}
\label{lenseq}
\end{equation}
Using these, the square of the distance of the source from the lens 
center is,
\begin{equation}
r_s^2=(r-f)^2+\Bigl({{\partial f}\over{\partial\theta}}\Bigr)^2,
\label{rs2}
\end{equation}
The determinant of the magnification matrix for our lensing potential is,
\begin{equation}
\det A={1\over r}\Bigl[(r-f)-{{\partial^2f}\over{\partial\theta^2}}\Bigr]
\label{detA}
\end{equation}

\begin{figure}
\epsscale{1}
\plotone{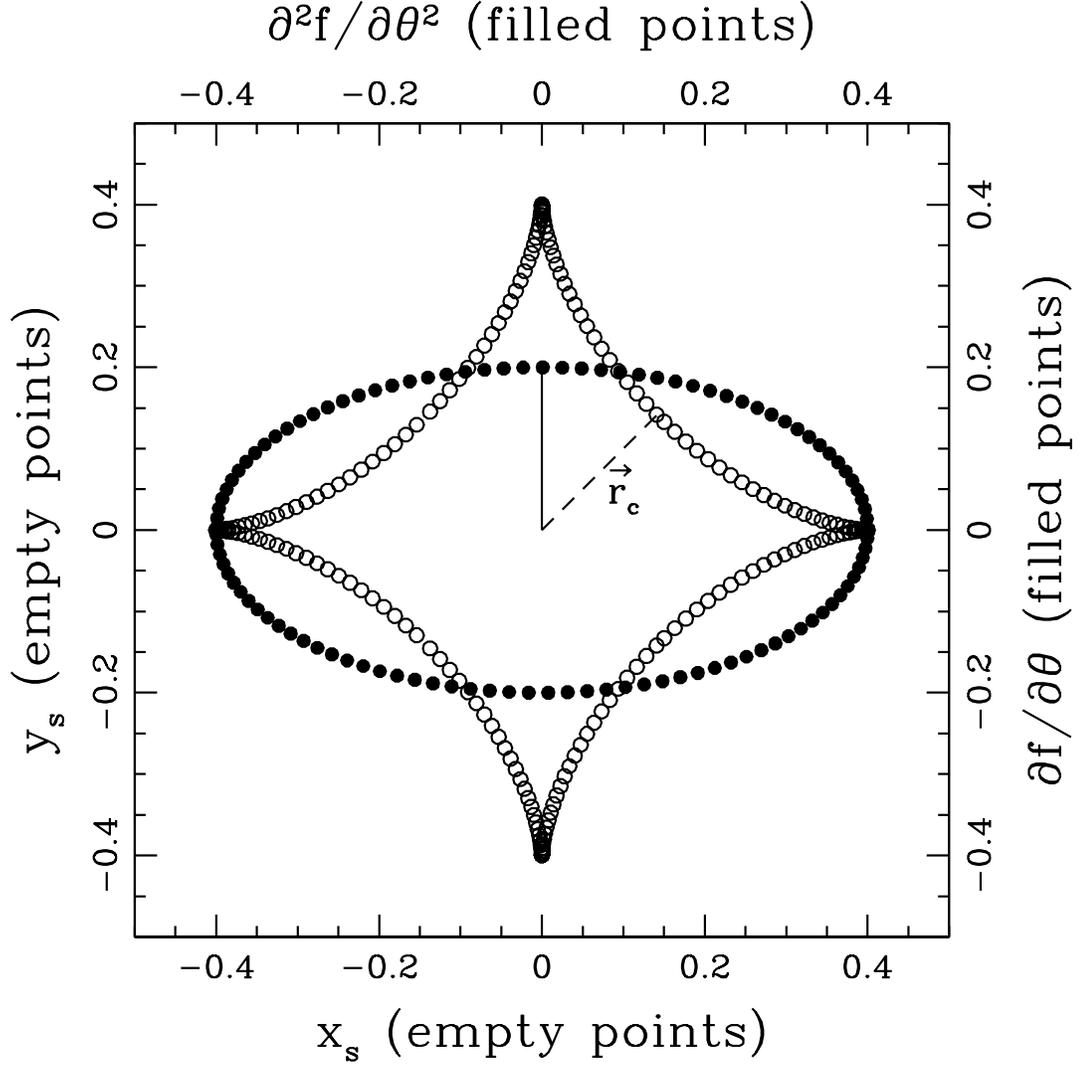}
\caption{The caustic has the usual diamond shape when plotted in the source
plane (empty points, and left and lower axes), but when plotted in the plane of
${{\partial f}\over{\partial\theta}}$ vs. ${{\partial^2f}\over{\partial\theta^2}}$
(filled points, and right and upper axes) it has an oval shape discussed in
Section~\ref{SISell}. The lensing potential used here is 
$\phi(r,\theta)=br(1+\gamma\cos 2\theta)$ with $b=1$ and $\gamma=0.1$. The solid
line segment represents the point of closest approach in the plane of
${{\partial f}\over{\partial\theta}}$ vs. ${{\partial^2f}\over{\partial\theta^2}}$,
while the dashed line is $\vec r_c$, in the source plane.}
\label{oval}
\end{figure}

For sources on the caustic, $\det A=0$, and so
$r-f={\partial^2f}/{\partial\theta^2}$. The caustic equation becomes
\begin{equation}
r_s^2=\Bigl({{\partial^2f}\over{\partial\theta^2}}\Bigr)^2
     +\Bigl({{\partial  f}\over{\partial\theta  }}\Bigr)^2.
\label{rs2caus}
\end{equation}
The two lensing equations, eq.\ref{lenseq} can then be rewritten as,
\begin{equation}
r_s\cos(\theta-\theta_s)={{\partial^2 f}\over{\partial\theta^2}},
\quad\quad\quad\quad
r_s\sin(\theta-\theta_s)={{\partial   f}\over{\partial\theta}}.
\label{lenseqcaus}
\end{equation}
Equations~\ref{rs2caus} and \ref{lenseqcaus} make it apparent that
the caustic is oval shaped in the plane defined by orthogonal axes equal
to the second and first derivatives of $f$ with respect to $\theta$,
respectively. The angle that specifies position in that plane is 
$(\theta-\theta_s)$. This oval is illustrated in Figure~\ref{oval}, with
filled points, and the right and upper axes. Note that this plane, where the 
caustic has an oval shape is not same as the source plane. For comparison,
the caustic in the source plane is also shown in Figure~\ref{oval}, with empty 
points, and the left and lower axes. In the source plane the caustic has the 
usual diamond shape.
 The point of closest approach belongs to the oval and is either on the
${{\partial^2 f}\over{\partial\theta^2}}$ axis, or on the
${{\partial f}\over{\partial\theta}}$ axis, i.e. either 
${{\partial f}\over{\partial\theta}}=0$, or
${{\partial^2 f}\over{\partial\theta^2}}=0$, respectively.

To proceed further we specify the form of $\phi$,
\begin{equation}
\phi(r,\theta)=br(1+\gamma\cos 2\theta),
\label{phi}
\end{equation}
where $b$ and $\gamma$ are constant for any given lens. This is the lensing 
potential of a singular isothermal sphere with an added elliptical perturbation, 
$\gamma$, which generates shear. If there were no shear, $b$ would be the 
Einstein ring radius of the SIS lens. This SIS+elliptical lens model 
is discussed, for example in \citet{dalal98}. For this lens,
\begin{equation}
{{\partial^2f}\over{\partial\theta^2}}=-4b\gamma\cos 2\theta,\quad\quad\quad\quad
{{\partial  f}\over{\partial\theta}}=-2b\gamma\sin 2\theta,
\label{derivs}
\end{equation}
which implies that the point of closest approach corresponds to 
${{\partial^2 f}\over{\partial\theta^2}}=0$. (This is shown as the solid line 
segment in Figure~\ref{oval}.) From the first of equations~\ref{lenseqcaus}, 
and restricting ourselves to the 1st and 4th quadrants
(the other two are redundant because of symmetry) we derive that 
$\theta-\theta_{c}=\pi/2$, $\theta=\pi/4$, and so $\theta_{c}=-\pi/4$. Here,
$\theta$ is the lens plane angle of only one of the images. $\theta_{c}$ is the 
angle of the point of the closest approach, $\vec r_c$ in the source plane, 
which is shown as the dashed line segment in Figure~\ref{oval} (left and lower
axes refer to the source plane).

According to our hypothesis all the points defining the bisector plot envelope
lie on a straight line. Therefore, having found its angle, namely $\theta_{c}$
we can now solve for the source positions themselves. To do this we use the
second of equations~\ref{lenseqcaus}. Squaring it, and using 
~$\sin^2\theta_{c}=\cos^2\theta_{c}=\frac{1}{2}$ we get,
\begin{equation}
\frac{1}{2}\Bigl[{{r_s}\over{2b\gamma}}\Bigr]^2(1-\sin 2\theta)
=\sin^2 2\theta
\label{quadratic}
\end{equation}
Here, $\theta$ refers to any one of four images, two minima and two 
saddles, and in fact this quadratic equation does have four solutions.
There are two solutions for $\sin 2\theta$ from the quadratic itself, and
each one of these gives two solutions because 
~$\cos 2\theta=\pm \sqrt{1-\sin^2 2\theta}$. 

The two images with $\sin 2\theta>0$ are in the 1st and 2nd quadrants, while 
the other two are in the 3rd and 4th. For each of these two pairs of images 
their $x$-coordinates place them equidistantly on either side of the $y$-axis. 
This implies that the angular distribution of the four images is symmetric about 
the $y$-axis. We can take advantage of this in determining how to sort these 4 
images in order of arrival time. First note that
images 2 and 3 are interchangeable; the same is true for images 1 and 4.
Images 2 and 3 are the ones that merge together when the
source is on the caustic. This happens for the largest possible $r_s$, i.e. 
$r_c=2b\gamma$. By considering various pairs of adjacent images in turn, one 
can show that of the 4 images the two that satisfy the merging criterion are
the ones with $\sin 2\theta=(\Delta+K)/2$, where $\Delta=\sqrt{K^2+4K}$, and 
$K=\frac{1}{2}[r_s/r_c]^2=\frac{1}{2}[r_s/2b\gamma]^2$. 
When the source is on the caustic $2\theta=-\pi/2$
for both of these. The other two images have to be 1 and 4. 
The angular separation between images 2 and 3 is then
\begin{equation}
\theta_{23}=
\pi/2-\tan^{-1}\Biggl[{{(\Delta+K)/2}\over{\sqrt{1-\frac{(\Delta+K)^2}{4}}}}\Biggr].
\label{th23}
\end{equation}
Similarly, the angular separation between images 1 and 4, which is always 
greater that $\pi/2$ is,
\begin{equation}
\theta_{14}=
\pi/2+\tan^{-1}\Biggl[{{(\Delta-K)/2}\over{\sqrt{1-\frac{(\Delta-K)^2}{4}}}}\Biggr].
\label{th14}
\end{equation}
Then, with some angle visualizing one arrives at the bisector angle difference as,
\begin{equation}
\beta_{12}-\beta_{34}=[2\pi-(\theta_{23}+\theta_{14})]/2
\label{bisd}
\end{equation}
This is what is plotted as the solid curve in Figure~\ref{bisector_twofold}, 
and subsequent similar figures.

\section{Real quads}\label{realquads}

Our quad lenses are taken from the CASTLeS data set \citep{castles}. 
We used all quads, except,
PMNJ0134-0931, whose lensing galaxy's position is ambiguous;
B0128+437, whose lens center is unknown;
SDSS1406+6126, which has partial data; and 
Q0047-2808, SDSS1029+2623, SDSS1402+6321 which have no data at all.
We also used two lenses that are not in CASTLeS: 
SDSS J125107.57+293540.5 \citep{k07}, and
HE1113-0641 \citep{bws07}.
Cluster lens SDSS J1004+4112, with QSO image separation of $\sim 15''$ was 
excluded because the images are formed by the central part of a galaxy cluster, 
not a galaxy. The source in B1933+503 is a double lobed radio source,
whose core and one of the lobes are each lensed into quads. These two quads 
were included as two separate lenses. This gives us a total of 26 quad lenses
listed in Tables~\ref{table1} and \ref{table2}. Lenses in Table~\ref{table1} have 
unambiguous arrival time ordering of images.

In some cross-like quads it is hard to know what the correct numbering of 
images should be. In the most ambiguous cases we can only be certain that images
1 and 2 should lie across from one another, and so should images 3 and 4. Using
this as the only rule gives us four distinct  
$(\beta_{12}-\beta_{34},~\theta_{23})$ pairs. However, two of these have 
unrealistically large $\theta_{23}$ values, generally in excess of $100^\circ$, 
and can therefore be discarded, leaving us with two possibilities for the 
$(\beta_{12}-\beta_{34},~\theta_{23})$ pair. There are 10 ambiguous lenses, 
and each one generates two lines in Table~\ref{table2}.

\begin{figure}
\epsscale{1}
\plotone{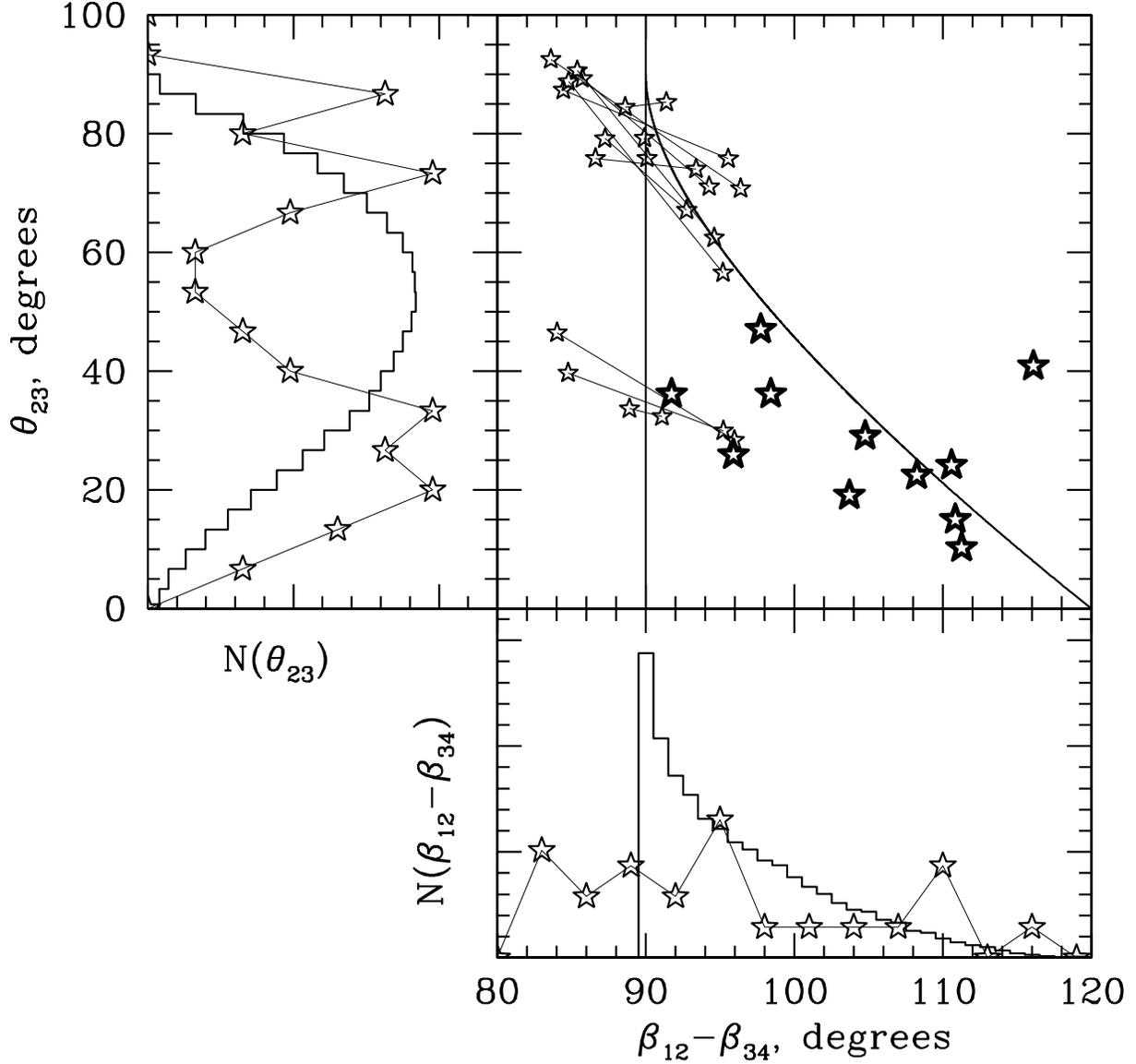}
\caption{Similar to Figure~\ref{bisector_twofold}. The main portion of the 
bisector plot shows the real lenses. Quads with unambiguous time ordering
(Table~\ref{table1}) are represented by the bold star symbols. The lenses
with ambiguous arrival time ordering (Table~\ref{table2}) are shown with
two star symbols each, connected by a thin line. The envelope curve
is given by eqs.~\ref{bisd} and \ref{th23}. In the two side panels, the 
solid line histograms represent two-fold symmetric lenses. The thin
line histograms delineated with star symbols represent the data. Each one
of the two bisector plane locations of the 'ambiguous' lenses was counted 
as $\frac{1}{2}$ in the histograms.}
\label{bisector_data}
\end{figure}

The quad data is shown in the bisector plot of Figure~\ref{bisector_data}.
The unambiguous arrival time order lenses are represented by bold star symbols.
Each one of the 10 ambiguous time order lenses is represented by two smaller
star symbols, connected by a thin line.

It is apparent from Figure~\ref{bisector_data} that the real quads are not 
drawn from the quad distribution generated by two-fold symmetric 
lenses. This is most clearly seen close to the 'apex' of the bisector
plot, near $(\beta_{12}-\beta_{34},~\theta_{23})=(90^\circ,90^\circ)$.
Here, nearly all star symbols lie outside of the apex outlined by two-fold
symmetric lenses. The lower portion of the two-fold symmetric lens bisector 
plot, roughly below $\theta_{23}\approx 60^\circ$ also appears to be inconsistent 
with the observed quad population: the latter are distributed more or less
evenly in the region below the envelope, whereas the density of small points
(from two-fold symmetric lenses) in Figure~\ref{bisector_twofold} increases 
sharply as one approaches the envelope from below. The final major difference
is that there is an apparent dearth of real lenses with $\theta_{23}\sim 50^\circ$,
which is not reproduced in the two-fold symmetric lenses.

The two solid line histograms in (the two side panels of) 
Figure~\ref{bisector_data} represent two-fold symmetric lenses, while the 
histogram delineated with star symbols are the quad data. The Kolmogorov-Smirnov 
(KS) test as applied to the $\theta_{23}$ distribution states
that the real quads could not have been drawn from the two-fold symmetric 
lenses at 95\% confidence level. The main reason for this is the lack of
real quads with $\theta_{23}$ around $50^\circ$, exactly where the two-fold
symmetric lenses predict most of the quads to lie.

The KS test applied to the $(\beta_{12}-\beta_{34})$ distribution is far less 
conclusive, but note that the KS test is not the optimal test here. In 
Section~\ref{twofold} we saw that no strictly two-fold symmetric lens can produce
$(\beta_{12}-\beta_{34})$ even a degree smaller than $90^\circ$. So the 
presence of real quads with $(\beta_{12}-\beta_{34})\sim 85^\circ$ rules 
out these lenses. We conclude that the population of real quads could not have 
been generated by two-fold symmetric galaxy lenses only. Many lensing galaxies
must have more complicated mass distributions.

In the next section we explore lenses with twisting isodens and lenses
with various degrees of substructure. That substructure may be important is
already suggested by HE0230. This lenses' image time ordering is unambiguous.
Its coordinates in the bisector plot of Figure~\ref{bisector_data} are at 
approximately ($116^\circ$,$41^\circ$), quite far above the envelope. 
According to the arguments of Section~\ref{twofold}, the lens mass distribution
must deviate strongly from two-fold symmetric.  And if fact, looking 
at the optical image of the lens (see CASTLeS) it is apparent that in 
addition to the main lensing galaxy there is a secondary galaxy, located
close to image 4. The spectroscopic data of \citep{e06} shows that the main 
lensing galaxy and the smaller secondary one are most probably members of a
galaxy group. A tentative conclusion, to be tested in the next section, is
that lens substructure in HE0230 and other lenses is responsible for the
disagreement between the bisector plots of two-fold symmetric lenses
and the real quad population.

\section{Mass distribution: Lenses lacking two-fold symmetry}\label{notwofold}

This is a large class of lens models, for example, lenses with twisting 
density contours, lenses with internal and external shear of different 
amplitudes and PAs, lenses with substructure, etc. Many real lenses belong in 
this vast category.

\begin{figure}
\epsscale{1}
\plotone{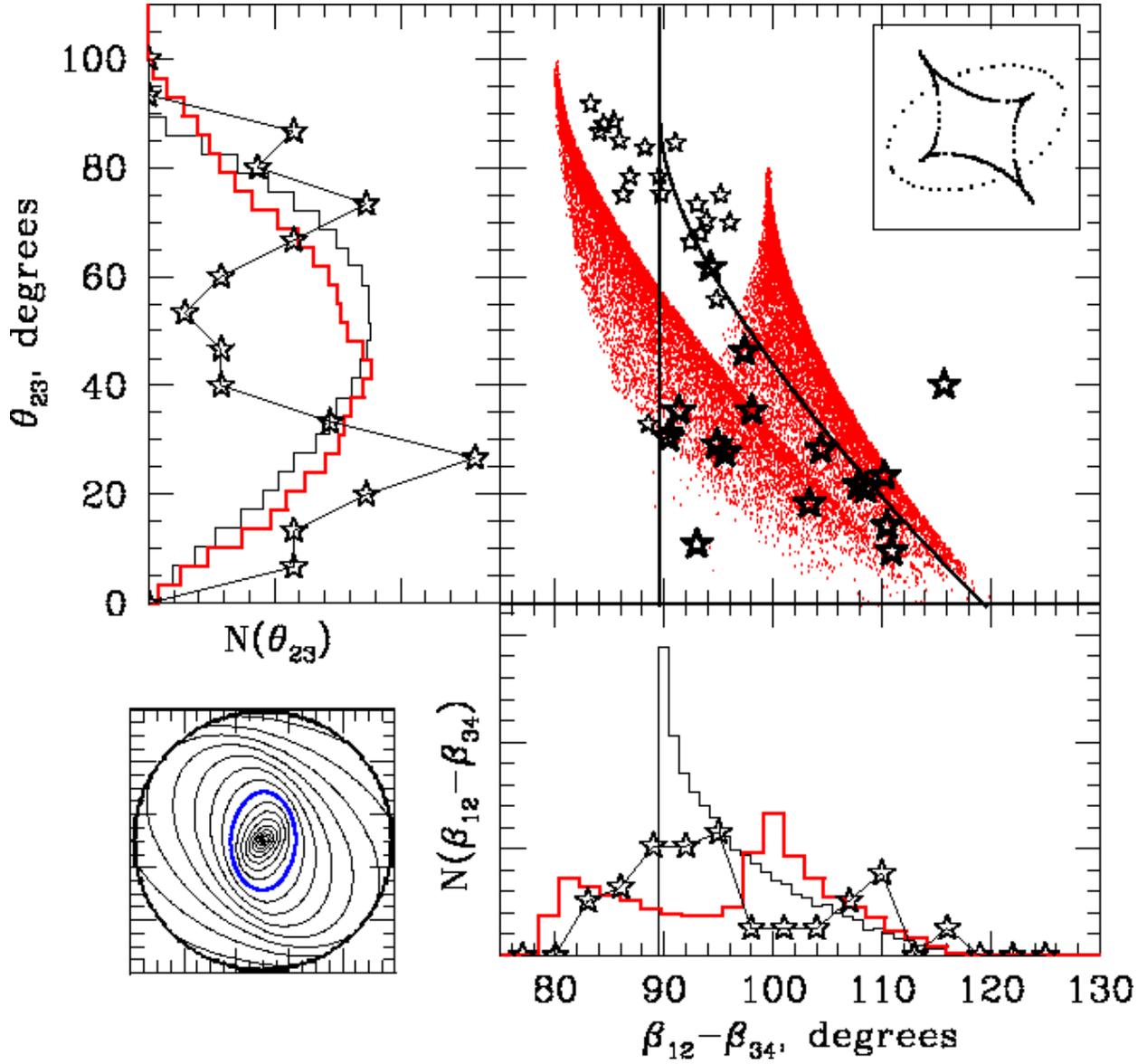}
\caption{Similar to Figures~\ref{bisector_twofold} and \ref{bisector_data}. 
The isodens of the lens
mass distribution are shown in the lower left inset. The thick (blue) contour
shows the critical surface mass density for lensing. The contours are spaced
logarithmically. (The lens the same as the one shown in Figure~\ref{fourpanels}.) 
The small (red) points in the main 
portion of the plot are the quads generated by this lens. The thick (red)
line histograms in the two side panels belong to this lens. The thin line
histograms are for two-fold symmetric lenses, shown here for comparison.
The inset in the upper right shows the source plane caustic.}
\label{bisectorRT}
\end{figure}

As a first example we take a synthetic galaxy lens with highly twisting 
isodens, the one shown in Figure~\ref{fourpanels}, and also in the
lower left inset in Figure~\ref{bisectorRT}. The thick (blue) contour has
the surface mass density equal to critical for lensing.
The main portion of the same
figure is the bisector plot. The single peak of 
Figure~\ref{bisector_twofold} has now split into two peaks. The upper right inset
in a plain line box shows the source plane caustics. In contrast to the
caustics of two-fold symmetric lenses, this diamond caustic is not
two-fold symmetric, for example, the lines connecting its opposite cusps are not 
perpendicular to each other. 

The left and bottom side panels of Figure~\ref{bisectorRT} show, in bold,
the $\theta_{23}$ and $\beta_{12}-\beta_{34}$ histograms for this lens.
As in the case of two-fold symmetric lenses, the real quad 
$\theta_{23}$ distribution does not match that of the synthetic lens with
twisting isodens, because the latter peaks, instead of dipping around $50^\circ$.

\begin{figure}
\epsscale{1}
\plotone{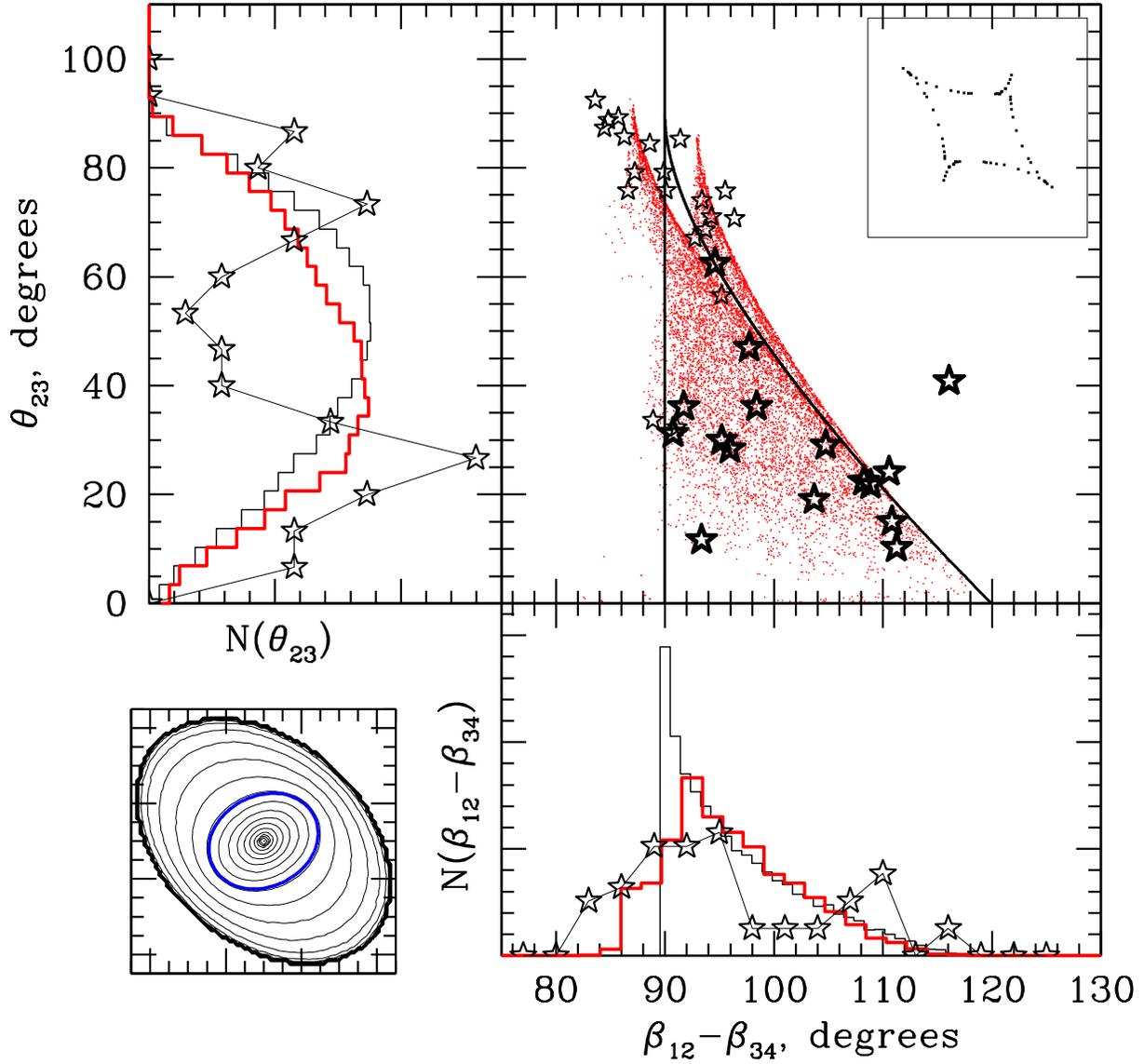}
\caption{Same as Figure~\ref{bisectorRT}, only for the lens shown in the 
lower left. See Section~\ref{notwofold} for details.}
\label{bisectorX1}
\end{figure}

The mass distribution of the Figure~\ref{bisectorRT} lens was not meant
to represent any real projected galaxy. Isoden twists in real galaxies result
from the projection of intrinsically triaxial galaxies with radially 
dependent axes ratios. To produce a more realistic isoden twisting we start
with a three dimensional mass distribution given by,
\begin{equation}
\rho(r)=(1+r/r_0)^{-2},\quad\quad\quad {\rm and} \quad
~r^2={x^2\over{a^2}}+{y^2\over{b^2/t}}+{z^2\over{c^2/t}},
\end{equation}
where $t$, a parameter proportional to $x$, governs the rate of change of
axis ratios with radius. We used $a:b:c=1:10:2$. Projecting this triaxial 
galaxy on to the plane of the 
sky using Euler angles $\phi=30^\circ$, $\theta=40^\circ$ and $\psi=100^\circ$ 
we get the mass map shown in the lower left inset of Figure~\ref{bisectorX1}.
The normalization of the mass distribution is such that the thick (blue) contour 
has the critical surface mass density for lensing. The difference in the PA of 
the inner and outer isodens is about $70^\circ$, consistent with what is 
observed for nearby galaxies \citep{l05}. For our purposes, this synthetic galaxy 
is a reasonable approximation for a typical projected triaxial galaxy.

Sampling the source plane caustic, shown in the upper right inset, using 
randomly placed sources we get the main panel of Figure~\ref{bisectorX1}.
This bisector plot looks similar to the one in Figure~\ref{bisectorRT},
only the separation of the peaks around $\beta_{12}-\beta_{34}=90^\circ$ is
smaller. In general, the spread of the peaks is directly related to the degree 
of isoden twisting in the lens. Just as in the case of Figure~\ref{bisectorRT},
this lens model, and by extension the population of realistic triaxial galaxies
cannot reproduce the bisector plot distribution of the real quads, primarily
because of the dearth of observed quads with $\theta_{23}$ near $50^\circ$.

Before we leave lenses with twisting isodens we note that elliptical lenses
with external shear whose axis does not coincide with the PA of the lens
produce bisector plots similar to the ones in Figures~\ref{bisectorRT} and
~\ref{bisectorX1}.

\begin{figure}
\epsscale{1}
\plotone{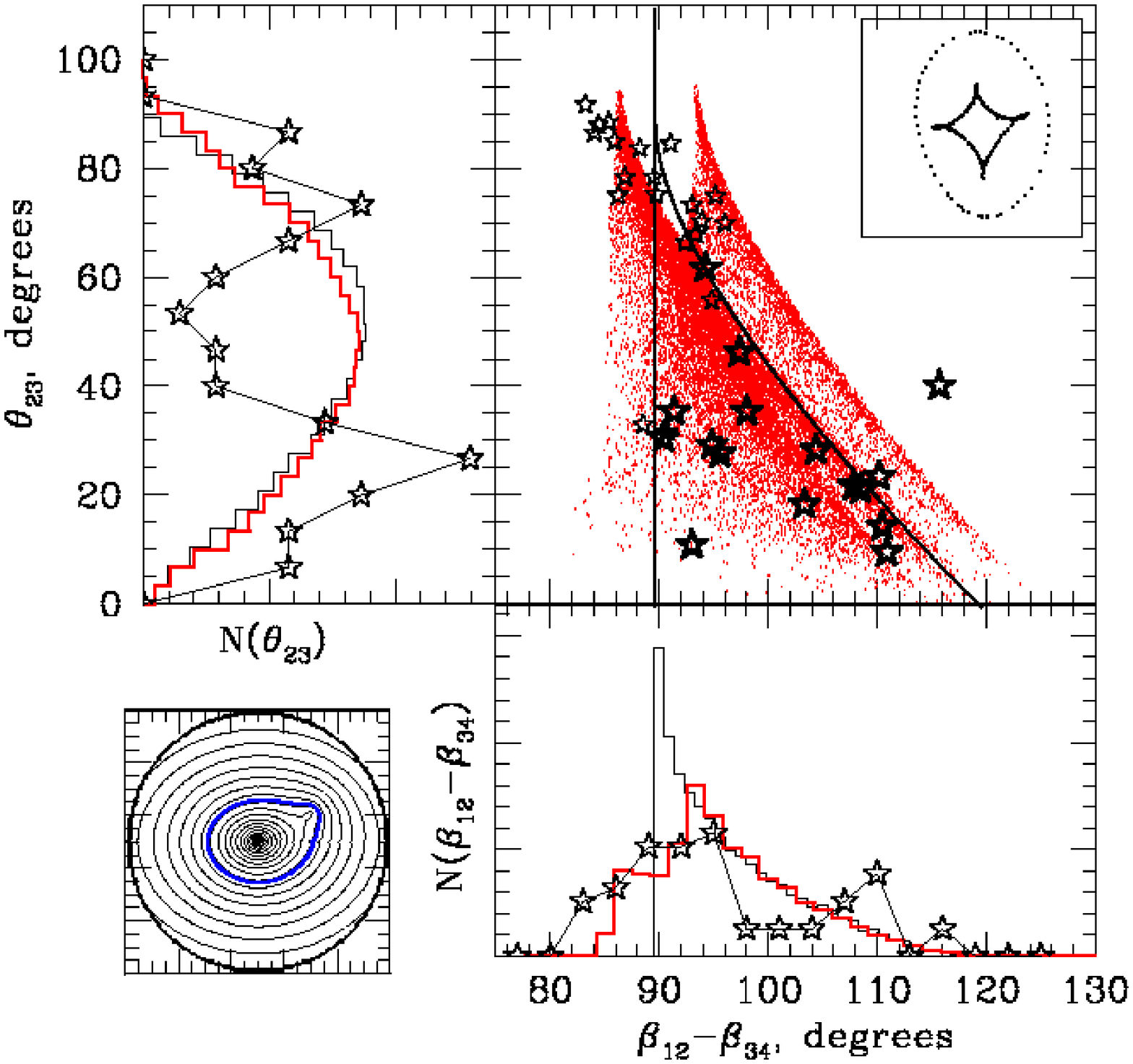}
\caption{Same as Figure~\ref{bisectorRT}, only for the lens shown in the 
lower left. The secondary galaxy comprises about 0.8\% of the total lensing mass.
The surface density profiles of the main and secondary galaxies are, respectively,
$\Sigma_m\propto \exp(-R/R_m)^{0.25}$ and $\Sigma_s\propto \exp(-R/R_s)$, and
$R_s/R_m=0.7$. See Section~\ref{notwofold} for details.}
\label{bisectorR5}
\end{figure}

\begin{figure}
\epsscale{1}
\plotone{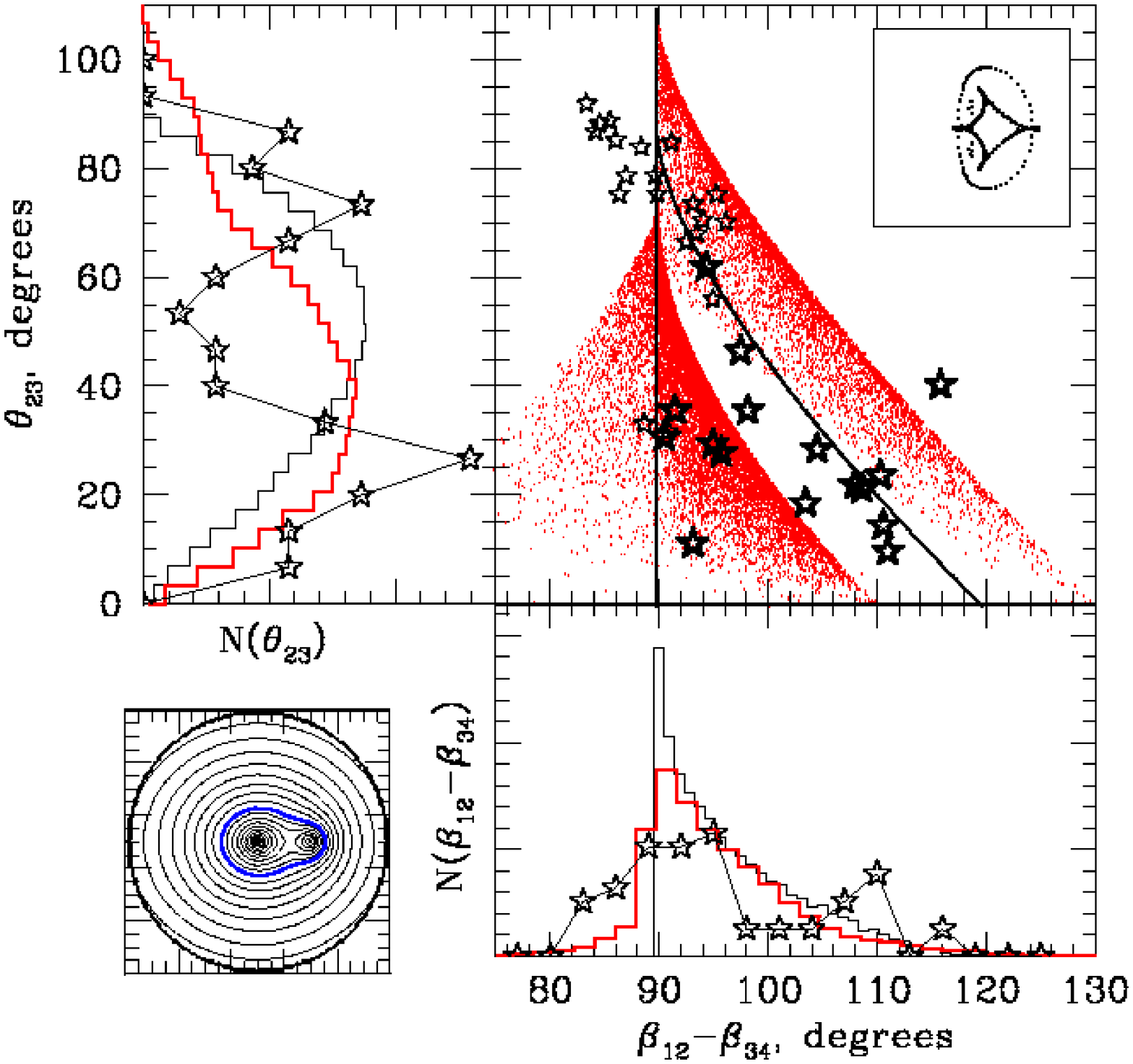}
\caption{Same as Figure~\ref{bisectorRT}, only for the lens shown in the 
lower left. The secondary galaxy comprises about 1.3\% of the total lensing mass.
The surface density profiles of the main and secondary galaxies are, respectively,
$\Sigma_m\propto \exp(-R/R_m)^{0.25}$ and $\Sigma_s\propto \exp(-R/R_s)$, and
$R_s/R_m=0.5$. See Section~\ref{notwofold} for details.}
\label{bisectorR7}
\end{figure}

Next, we turn to lenses with substructure lumps, like secondary or satellite
galaxies located close the primary lens galaxy. Our goal here is to consider a few
representative substructure types. A systematic exploration of the substructure
and what matches observations best will be done in a later paper. 
Figures~\ref{bisectorR5} and \ref{bisectorR7} show results for lenses with one
subclump each. In the first case, Figure~\ref{bisectorR5}, the subclump 
represents a small perturbation to the lens, so the caustic is only slightly 
distorted from its two-fold symmetric diamond shape. Because the lens is now
more complex, the bisector plot is also more complex. However, the 
$\theta_{23}$ distribution still does not look like that of the real quads.

In the second case, Figure~\ref{bisectorR7}, the subclump 
is compact and relatively more massive. Here, the lens' $\theta_{23}$ 
distribution (left side panel) looks quantitatively different from all the ones 
we have considered so far; it is not a single peaked distribution, centered at
about $55^\circ$. The main peak has moved to $40^\circ$, and there is an
incipient second peak close to $\theta_{23}=90^\circ$. Furthermore, the bisector
plot points are beginning to extend far above the envelope, almost reaching
HE0230, the 'outlier' at ($116^\circ,~41^\circ$). Perhaps it is not surprising
that this lens model (almost) reproduces HE0230; the lens model contains a major 
secondary perturber, just as the real lens in the HE0230 system.

\begin{figure}
\epsscale{1}
\plotone{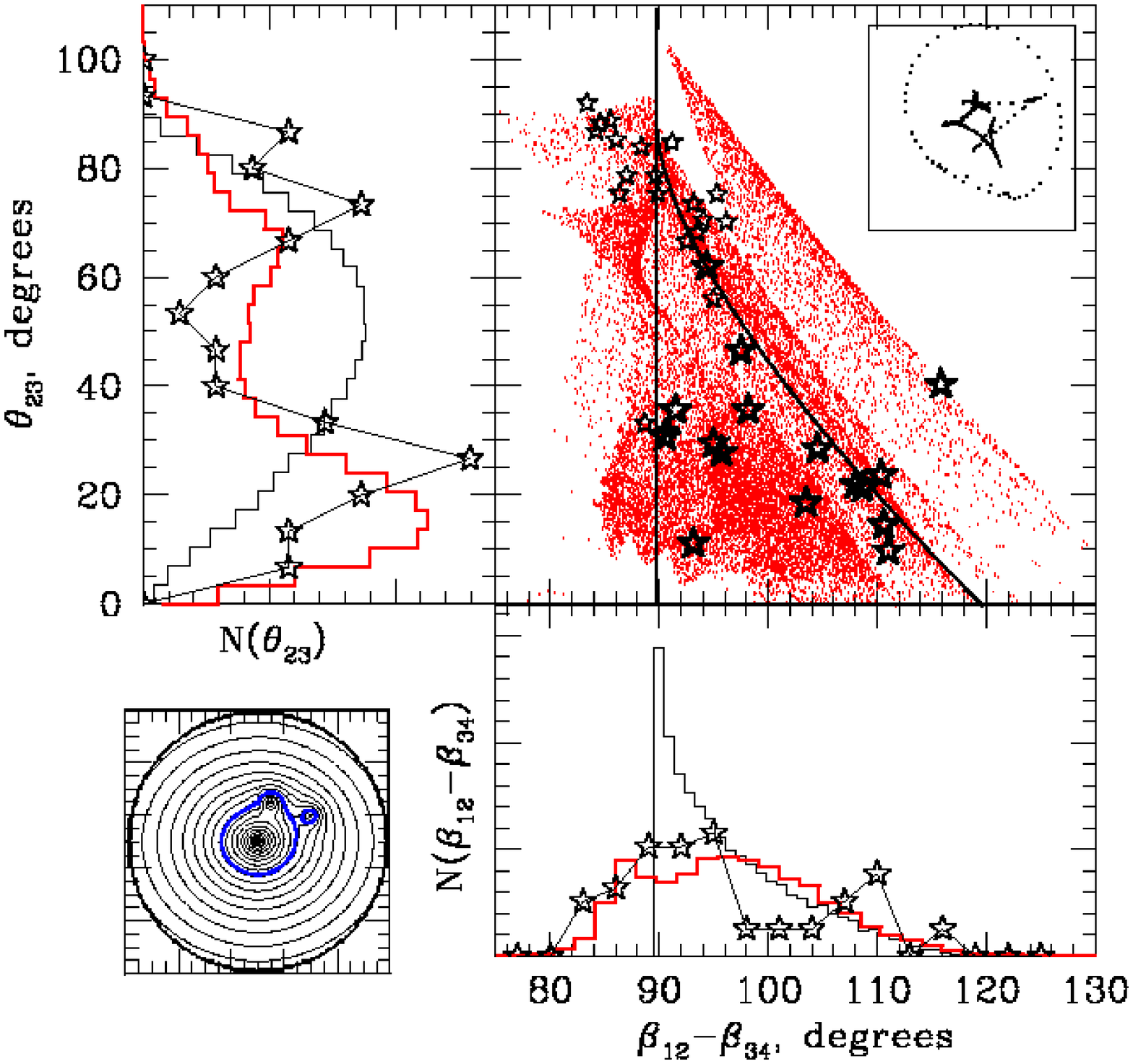}
\caption{Same as Figure~\ref{bisectorRT}, only for the lens shown in the 
lower left. The two secondary galaxies together comprises about 0.4\% of the total 
lensing mass. The surface density profiles of the main and the two secondary 
galaxies are, respectively, $\Sigma_m\propto \exp(-R/R_m)^{0.25}$ and 
$\Sigma_s\propto \exp(-R/R_s)$, and $R_s/R_m=0.2$. 
See Section~\ref{notwofold} for details.}
\label{bisectorR6}
\end{figure}

Figure~\ref{bisectorR6} shows the results for a lens with two substructure
clumps. The caustic bears no resemblance to a diamond shape, and the bisector
plot distribution is very complex. This lens model reproduces, at least 
qualitatively, major features of the observed quad distribution in the bisector
plane. Note that we did not aim to do so; no effort was put 
into matching the observed distribution in any detail. The dearth of quads at 
$\theta_{23}\sim 50^\circ$ is present in the synthetic lens, and the distribution
of points in the bisector plane extends all the way to HE0230, something that even
the lens of Figure~\ref{bisectorR7} could not do.

Figures~\ref{bisectorX1}--\ref{bisectorR6} are meant only as qualitative 
guides to different types of non two-fold symmetric lenses. Based on these
we tentatively conclude that the real population of quad lenses requires 
lumpy substructure; features like twisting isodens and external shear are
not enough. However, a thorough exploration of the parameter space of lenses 
is needed to make robust conclusions. This will be the subject of a later paper.

\section{Real doubles}\label{doub}

As the source of a quad system moves further away from the lens center
images 2 and 3 move closer to each other, and closer to the critical
line, and eventually disappear, transforming the lens into a double.
As a quad turns into a double, $\theta_{23}=0$ and the remaining images, 
1 and 4, become the two images of a double. Figure~\ref{bisector_twofold}
tells us that the largest bisector difference in a quad is $120^\circ$. 
Combining this with eq.~\ref{bisd} tells us that ``newly formed'' doubles 
should have $(2\pi-\theta_{14})/2=120^\circ$, i.e. their image separation
should be at least $\theta=120^\circ$. So, there should be no doubles with 
image separation $<120^\circ$. If the lens is not two-fold symmetric this 
limiting angle can change a little.

\begin{figure}
\epsscale{1}
\plotone{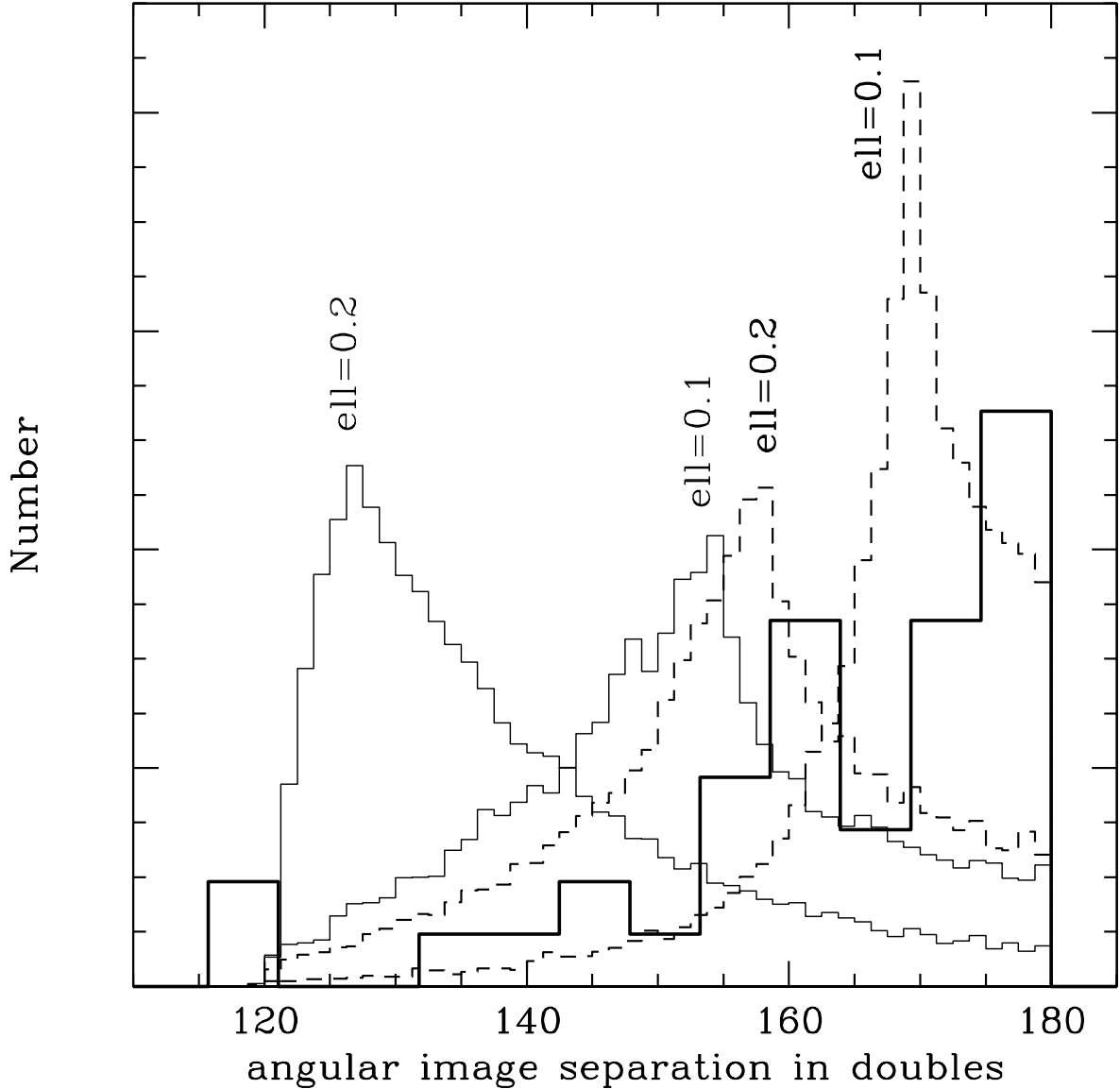}
\caption{The distribution of angles between two images of a double. The thick
line histogram shows 39 real doubles. The other four histograms represent 
synthetic lenses. The two thin solid line histograms correspond to galaxy 
lenses with projected density profiles $\propto \exp(-R^{0.25})$. The two 
dashed histograms represent ``isothermal'', $\propto R^{-1}$ lenses with a 
small core. The ellipticities, $e=0.1$ (axis ratio=0.82) and $e=0.2$ 
(axis ratio=0.67) of the lenses are labeled in the plot.
See Section~\ref{doub} for details.}
\label{doubles}
\end{figure}

Because doubles have only two images there is no such thing as a bisector 
plot for doubles, however, one can make a plot equivalent to the bottom 
panels of Figures~\ref{bisector_twofold}-\ref{bisectorR6}. This is shown in 
Figure~\ref{doubles}. The thick solid line histograms the angle between the 
two images of 39 doubles taken from CASTLeS. As expected, the angle between
the two images generally stays above $120^\circ$.

The other four histograms in Figure~\ref{doubles} represent synthetic lenses.
The two thin solid line histograms correspond to galaxy lenses whose
projected density profile is proportional to $\exp(-R^{0.25})$.
The two dashed histograms represent ``isothermal'' lenses with a small core;
outside the core the projected density scales as $R^{-1}$. Each one of these 
density profiles was given two, constant in radius, ellipticities: 
$\epsilon=0.1$ (axis ratio, $r=0.82$) and $\epsilon=0.2$ (axis ratio, $r=0.67$). 
Each one of the two
shallower lenses were given the same ellipticities. The ellipticities are
labeled in the plot. All four synthetic lenses are two-fold symmetric,
but, in contrast to the quads, the distributions of these lenses in the 
equivalent $\beta_{12}-\beta_{34}$ are different.

The conclusion we draw is that the distribution of doubles in angles is a 
more complex function of the galaxy lens parameter that is the case for quads.
A more detailed exploration of the doubles distribution in angles, perhaps
coupled to the analysis of the quads, will be a subject of another paper.

\section{Summary and Conclusions}

We introduce a novel way of analyzing the projected mass distribution in galaxy 
lenses that relies on the angular distribution of images in quads and doubles 
around the lens center. If the images of a quad are numbered in order of arrival, 
as $\theta_1$, through $\theta_4$, and $\theta_{ij}$ is the angle between images 
$i$ and $j$ then we define the bisector plane whose axes are linear combinations 
of $\theta_{23}$ and $\theta_{14}$. We show empirically that all two-fold symmetric
lenses with convex isodensity contours are identical when considered in the 
bisector plane. We derive an analytical expression for the boundary envelope of
the allowed region, for a specific type of lens. These results concerning the
invariance of the bisector plane for two-fold symmetric lenses is one of the main 
findings of the paper. It means, for example, that from the point of view of 
$\theta_{23}$ and $\theta_{14}$ of quads, a Pseudo Isothermal Elliptical Mass 
Distribution is identical to a circular lens, with any density profile plus an 
external shear. 

This invariance of the bisector planes of two-fold symmetric lenses can be used 
to examine the structure of the real galaxy lenses. We conclude that the 
observed quad population was not produced by two-fold symmetric lenses.

We also look at three realistic types of non two-fold symmetric mass distributions,
(1) galaxies with twisting isodensity contours, and elliptical galaxies with
external shear axis, (2) galaxies with single substructure clumps, and
(3) galaxies with two substructure clumps. It appears that only the last type
of lenses is able to reproduce the real quad population. This of course does not
mean that all galaxies with observed quads are of type (3), but it does
suggest that kpc-scale substructure is a common feature in galaxy lenses.

To confirm and quantify this conclusion a much more detailed exploration of the 
parameter space of non two-fold symmetric lenses is needed. Such a study should 
also include potential sources of bias in the quads. For example,
in this paper we have assumed that the real lenses represent a random 
sampling of the relevant region in the source plane; in other words, all
sources have the same weights. This means that we have neglected magnification 
bias, which makes sources at certain source plane locations more magnified, 
and hence more likely to enter a magnitude limited sample. The bias is probably 
negligible for quads, since they are already highly magnified; after all, 
quads are closely related to Einstein rings. It is unlikely that there is a 
missing population of faint quads. However, the magnification bias could be an 
issue for the doubles, and will need to be taken into account in future work.

Two final notes are in order. First, the lumpy substructure we refer to here
is different from that searched for using image flux anomalies, e.g. \citet{m04}.
In the latter case substructure lumps are small, and have to lie close to the
line of sight to the images. Our substructure lumps are larger, kpc-sized, more 
extended and can live anywhere within the central several kpc of the galaxy lens 
center. Second, the varied and complex lumpy substructure that our analysis implies 
the lenses should have argues strongly for using non-parametric, or 
semi-parametric modeling techniques.

\acknowledgements
This work was supported in part by NSF grant AST 03-07604, and 
HST-AR-10985.01-A.

\appendix
\section{Estimating the PA of the lens' major axis}\label{estimatingPA}

\begin{figure}
\epsscale{1}
\plotone{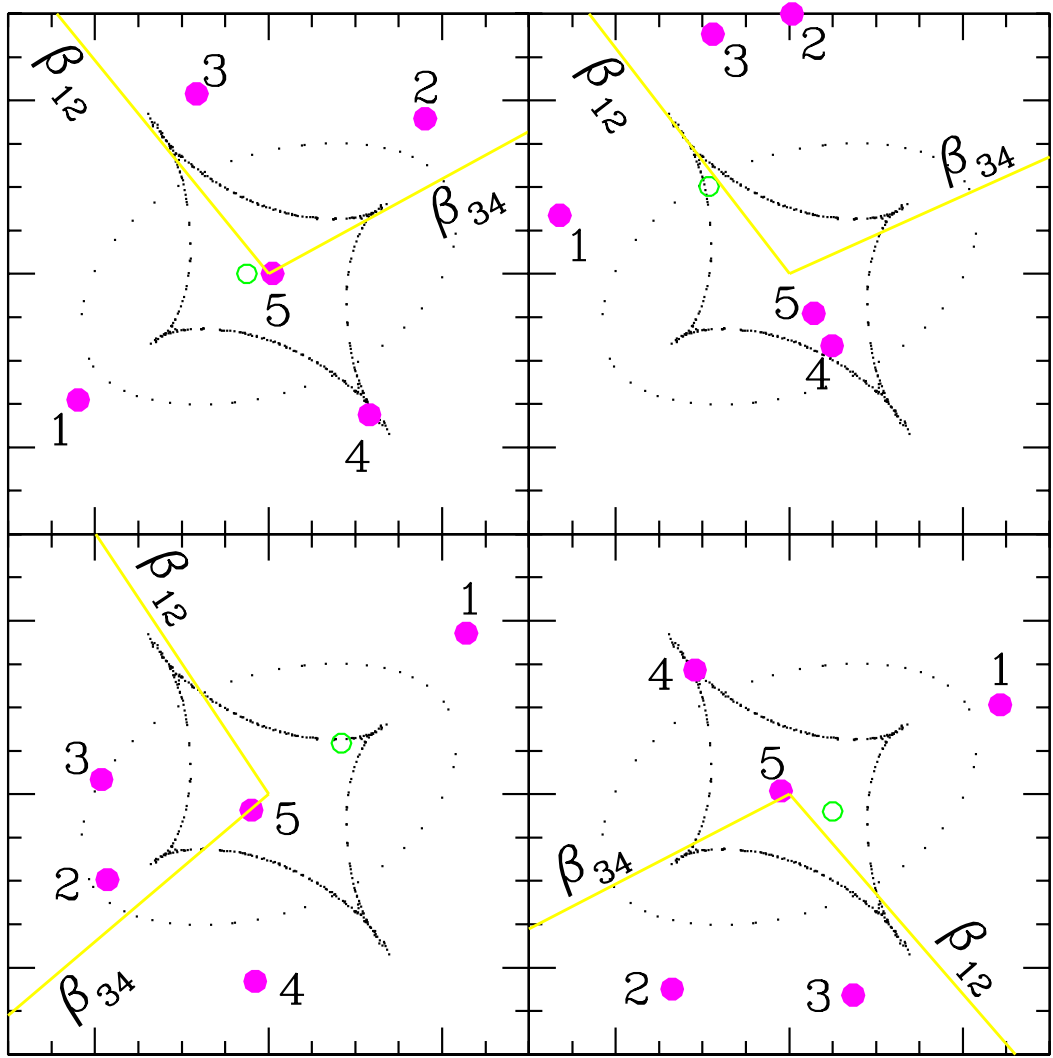}
\caption{All four panels show the caustics of the same lens as in 
Figure~\ref{fourpanels}, but four different source positions, empty (green) 
circle. Solid (magenta) dots numbered 1-5 are the {\em scaled down} positions 
of images, i.e. angles from the lens center are preserved, but distances are 
not. The two bisector rays, shown by straight (yellow) lines are labeled in 
each of the panels.}
\label{bisec}
\end{figure}

A given lens system can produce a variety of image configurations, depending
on the location of the source. The four panels of Figure~\ref{bisec} show the 
same lens galaxy with four different source positions. As the source location 
changes the angular positions of the images, and their angular separation
also change considerably.  However, the axis containing the bisector rays 
$\beta_{12}$ and $\beta_{34}$ change very little (i.e. $\beta_{12}$ and 
$\beta_{34}$ modulo $\pi$). Furthermore, the axes containing $\beta_{12}$ and 
$\beta_{34}$ coincide with the major and minor axes of the diamond caustic,
respectively, to within $\sim 10^\circ$.

\begin{figure}
\epsscale{0.9}
\plotone{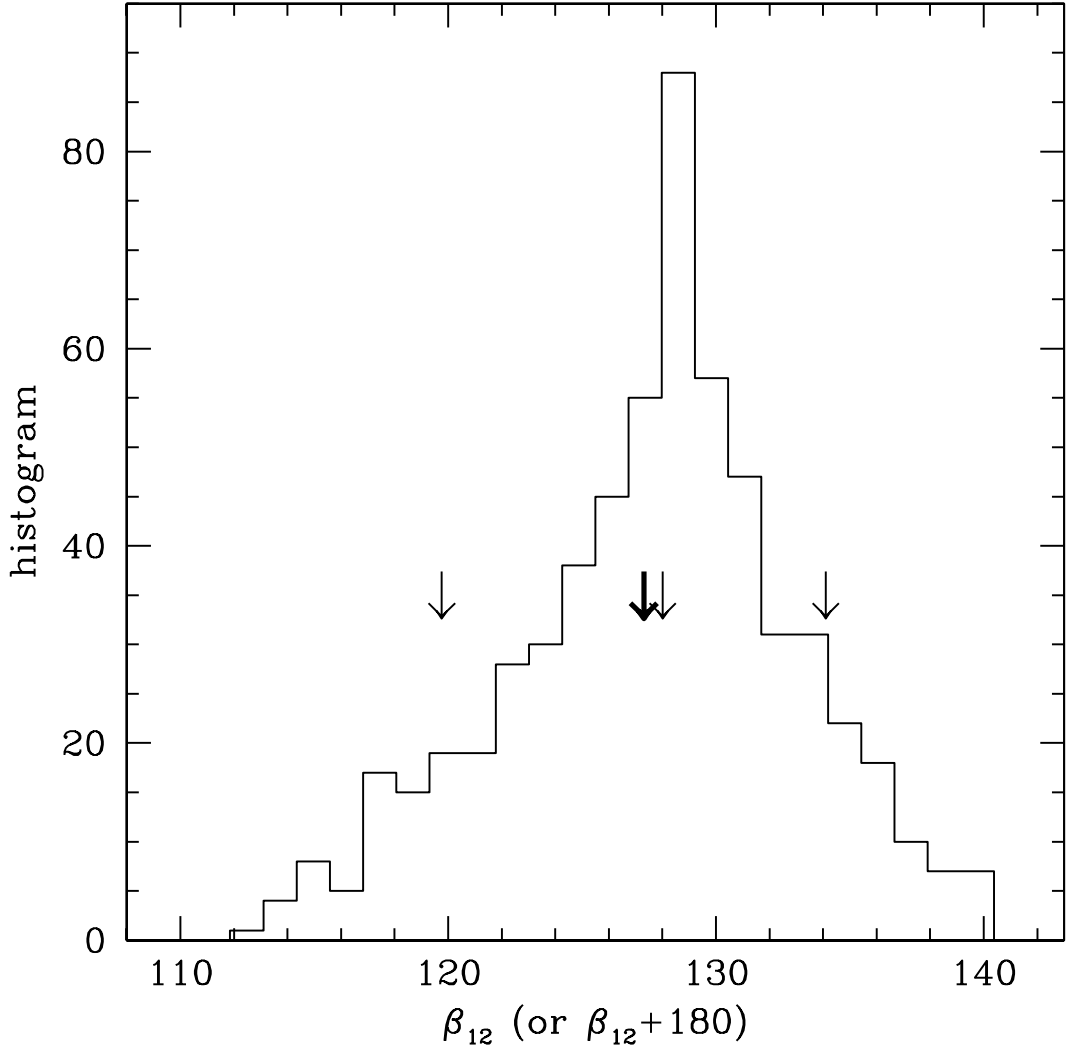}
\caption{The histogram of the PA of the axis containing the bisector ray 
$\beta_{12}$ for the lens shown in Figures~\ref{fourpanels} and \ref{bisec}.
Each source position contributes one value to the histogram.
The thick arrow indicates the {\it actual} PA of the major axis of the 
diamond caustic. The actual PA is very nearly the same as the mode and 
the median (central thin arrow) of the $\beta_{12}$ distribution.
The other two thin arrows mark the 10th and 90th percentiles. The plot 
illustrates that the PA of the axis containing the bisector ray $\beta_{12}$ 
coincides, to within a few percent, with the major axis of the diamond caustic,
and hence the major axis of the lens' mass ellipticity in the image circle.
See Section~\ref{estimatingPA} for details.}
\label{angles5} 
\end{figure}

Figure~\ref{angles5} this observation; it histograms the angle containing 
the $\beta_{12}$. Because this can be either $\beta_{12}$ or $\beta_{12}+\pi$
the full possible range is $\pi$. Each point contributing to the histogram 
represents a random source position. 
The thick down-arrow indicates the actual PA of the major axis of the 
diamond caustic. This is very nearly the same as the mode (peak of the 
histogram) and the median (central thin arrow) of the $\beta_{12}$ 
distribution. The other two thin arrows mark the 10th and 90th percentiles.
The half-width of the $\beta_{12}$ 
distribution is about $7^\circ$, i.e. the axis containing $\beta_{12}$ is
$\pm 7^\circ$ from the true PA of the major axis of the diamond caustic.
This means that for an unknown position of the source, measuring the axis
of $\beta_{12}$ and equating it to the axis containing the cusps of the caustic
will typically result in a $7^\circ$ error, or, fractional error of about 4\%.
 
In the absence of strong external shear
the direction of the major axis of the diamond caustic is aligned with 
the major axis of the mass distribution in the ring of the images. 
Therefore the PA 
of the axis containing bisector $\beta_{12}$ is aligned with the major axis of 
the mass distribution at the radius of the images.
In ~\citet{sw03} we noted that the direction of the dominant shear or 
ellipticity in a lens can be determined from the images:  The images lie on 
an eccentric ellipse whose major axis is perpendicular to the major axis of 
the dominant shear (whether it is internal or external). Here we suggest a 
more precise measure of the direction of the mass ellipticity, namely,
the axis containing $\beta_{12}$.

\clearpage

\clearpage

\begin{deluxetable}{rrl}
\tablewidth{0pc}
\tablecaption{Lens with unambiguous arrival time ordering}
\tablehead{
\colhead {$\beta_{12}-\beta_{34}$} & {$\theta_{23}$} & Lens name }
\startdata
  103.71  &  19.08  &  MG0414+0534 \\
  110.59  &  24.13  &  PG1115+080  \\
  116.09  &  40.85  &  HE0230-2130  \\
   97.74  &  47.00  &  SDSS0924+0219  \\
  111.27  &  10.25  &  B0712+472  \\
  108.25  &  22.46  &  HS0810+2554  \\
  110.84  &  15.01  &  B1933+503 (lobe) \\
  104.77  &  29.05  &  WFI2026-4536 \\
   98.41  &  36.14  &  WFI2033-4723 \\
   91.74  &  36.11  &  B1608+656  \\
   95.23  &  29.95  &  RXJ0911+0551 \\
   90.76  &  31.25  &  SDSSJ125107 \\
  108.89  &  21.79  &  B1555+375 \\
   94.61  &  62.47  &  SDSS1138+0314 \\
   95.97  &  28.37  &  B1422+231 \\
   93.38  &  11.66  &  B2045+265 \\
\enddata
\label{table1}
\end{deluxetable}

\clearpage

\begin{deluxetable}{rrl}
\tablewidth{0pc}
\tablecaption{Lens with ambiguous arrival time ordering}
\tablehead{
\colhead {$\beta_{12}-\beta_{34}$} & {$\theta_{23}$} & Lens name }
\startdata
   91.10  &  32.34  &  RXJ1131-1231 \\
   88.90  &  33.67  &  `` \\
   93.39  &  74.05  &  HST12531-2914 \\
   86.61  &  75.83  &  `` \\
   84.80  &  88.77  &  B1933+503 (core) \\
   95.20  &  56.57  &  `` \\
   91.39  &  85.29  &  SDSS1011+0143 \\
   88.61  &  84.44  &  `` \\
   85.73  &  89.28  &  H1413+417 \\
   94.27  &  71.07  &  `` \\
   95.55  &  75.77  &  HST14176+5226 \\
   84.45  &  87.34  &  " \\
   96.39  &  70.73  &  HST14113+5211 \\
   83.61  &  92.51  &  `` \\
   87.24  &  79.19  &  Q2237+030 \\
   92.76  &  67.11  &  `` \\
   90.09  &  75.88  &  HE0435-1223 \\
   89.91  &  79.25  &  `` \\
   93.72  &  68.65  &  HE1113-0641 \\
   86.28  &  85.76  &  `` \\
\enddata
\label{table2}
\end{deluxetable}

\end{document}